\documentclass[runningheads]{llncs}

\usepackage{subcaption}
\usepackage[utf8]{inputenc}
\usepackage{setspace}
\usepackage{afterpage}  
\usepackage{subcaption}
\usepackage{xcolor}
\usepackage{url}
\usepackage{wrapfig}
\usepackage{textcomp} 
\usepackage{comment}
\usepackage[font={scriptsize}]{caption}
\usepackage{ifthen}
\usepackage{amssymb}
\usepackage{amsmath}
\usepackage{graphicx}
\usepackage[]{easy-todo} 
\usepackage{soul}
\usepackage{multirow}
\usepackage{booktabs}
\usepackage{siunitx}
\usepackage{balance}
\usepackage{rotating}
\usepackage{tabularx}
\usepackage{float}
\usepackage{amssymb}
\usepackage{makecell}
\usepackage{expl3}
\usepackage{tablefootnote}
\usepackage{pbox}
\usepackage{mwe}
\usepackage{lipsum,etoolbox}
\usepackage{array}
\usepackage{lipsum}
\usepackage{courier}
\usepackage{listings}
\usepackage[]{easy-todo}
\usepackage{longtable}
\usepackage{pdfpages}
\usepackage{array}
\usepackage{bbding}
\usepackage{hyperref}

\usepackage{tikz}
\usepackage{lmodern}
\usepackage{nicematrix}
\usepackage{paralist}

\newcolumntype{P}[1]{>{\raggedleft\arraybackslash}p{#1}}
\newcolumntype{L}[1]{>{\raggedright\arraybackslash}p{#1}}
\usepackage{floatpag}
  \usepackage{graphicx,scalerel}

\makeatletter
\def\els@aparagraph[#1]#2{\elsparagraph[#1]{#2}}
\def\els@bparagraph#1{\elsparagraph*{#1}}
\makeatother

\usetikzlibrary{arrows.meta,fit,positioning,shapes,shapes.symbols,shapes.geometric,bpmn}
\usepackage{lmodern}
\usepackage{nicematrix,booktabs}

\usepackage{algorithm}
\usepackage{algorithmicx}
\usepackage[misc]{ifsym}
\usepackage{hyperref}
\usepackage{caption}

\newcommand{\PlusP}{\mathord{\begin{tikzpicture}[baseline=0ex, line width=1.5, scale=0.13]
\draw (1,0) -- (1,2);
\draw (0,1) -- (2,1);
\end{tikzpicture}}}
\newcommand{\Xor}{\mathord{\begin{tikzpicture}[baseline=0ex, line width=1.25, scale=0.08]
\draw (0,0) -- (2,2);
\draw (0,2) -- (2,0);
\end{tikzpicture}}}

\begin{document}
\title{Pragmos: A Process Agentic Modeling System}
\titlerunning{}
%
\author{Pedro-Aarón Hernández-Ávalos \and Luciano Garc\'{\i}a-Ba\~nuelos\textsuperscript{(}\Envelope\textsuperscript{)}}
\authorrunning{P.A. Hernández and L. Garc\'{\i}a-Ba\~nuelos}
%
\institute{Tecnologico de Monterrey, Mexico\\
\email{\{pedroaaron.hernandez,luciano.garcia\}@tec.mx} }
\maketitle              
\begin{abstract}
The advent of Large Language Models (LLMs) has significantly transformed tasks across Software Engineering. In the context of Business Process Management, LLMs are now being explored as tools to derive process models directly from textual descriptions. Existing approaches range from chatbot-driven systems that assist with iterative, text-based modeling to fully automated end-to-end modeling assistants.

However, we argue that process modeling is inherently complex and cannot be effectively addressed through black-box solutions. Instead, we envision modeling as an open-ended conversational activity, best supported by an interactive, iterative process involving both humans and LLM. In our approach, the modeling task is decomposed into smaller, manageable steps. Each step results in intermediate artifacts and explicitly documents the rationale behind each modeling decision. During this process, we incrementally uncover simple behavioral relations that guide the construction of the model.
Given the current limitations of LLMs in reasoning about complex dependencies, we complement them with specialized tools developed in the field to structure process models based on behavioral relations. This hybrid approach enables the generation of sound, yet comprehensible models that evolve through transparent and explainable steps.
%
In this paper, we present our research agenda and introduce Pragmos, a prototype system that operationalizes this vision. Pragmos demonstrates how LLMs can collaborate with human users as both domain and modeling experts to co-create evolving process models through a structured and explainable workflow.
    

\end{abstract}

\section{Introduction}
\label{sec:intro}
Process models are fundamental to business process management. They provide a representation of how an organization functions, enabling analysts to understand, analyze, track, and ultimately improve organizational practices~\cite{DumasRMR18}. However, creating accurate and useful process models is a complex undertaking. Business analysts must skillfully transform specialized knowledge, internal regulations, and operational constraints into clear, unambiguous representations –whether diagrams or detailed documentation– that faithfully reflect the actual work being performed.

Evidently, modeling implies a substantial cognitive load, as it requires a deep procedural insight, an understanding of the modeling conventions, and the ability to detect downstream implications such as deadlocks, bottlenecks, or even regulatory violations. On the one hand, one must consider the load connected with understanding the subtleties of work practices in the domain and knowledge behind the organization being targeted. On the other hand, empirical studies have shown that business analysts face an additional cognitive load connected with the inherent complexity induced by the constructs offered by the notation used to express the process model. For instance, \cite{PinggeraSFWZWRM15} shows that in addition to the time spent to reasoning about the domain, modelers are often involved in iterative editing, rearranging modeling elements, which suggests a substantial effort devoted to complying with syntactical constraints. Moreover, \cite{BockC18} observed that modeling sessions included numerous syntactical errors (e.g., incorrect use of gateways, among others), which can also be associated with the inherent complexity of notation.

The emergence of large language models (LLMs) has opened a new frontier for alleviating this burden. LLMs have shown remarkable success in tasks such as natural language understanding, processing, and generation, making them powerful tools for text generation, summarization, translation and even software artifact generation. LLMs are ML models trained with massive corpora which most likely include textual description of business processes, domain documentation and potentially formal process representations. Based on such observation, a bulk of research has explored the use of solely LLM prompting or some hybrid approach, e.g., using auxiliary process mining tools, to enable the construction of assistants capable of generating process model fragments, decision tables, and even complete models for simple textual scenarios. Indeed, delegating modeling tasks to LLM-based tools free analysts to focus on strategic reasoning, validation, and iterative refinement. Hence, the emergence of these LLM‑based tools opens a co-design loop, due to the conversational nature of their operation, in which stakeholders and analysts can iteratively clarify requirements at the time the model evolves. However, the work known by the authors considers these assistants as black-boxes, or relies on highly technical internal representations.

In this work, we advocate for a novel approach where the outcomes of the several steps of modeling are always intuitive artifacts (e.g., set of execution paths, sets of activities involved in a loop, etc.) allowing analysts to question/refine the intermediate results and having full control thereof. Our approach relies on the use methods developed for structuring process models, which provides a robust theory to support the synthesis of sound process models, while still relying on the intuitive artifacts we mentioned before. Our ideas have been implemented as a prototype called Pragmos, which demonstrates the feasibility of the method. We further evaluated the prototype with a well-known dataset, showing encouraging results.

The remainder of the paper is organized as follows. In Section~\ref{sec:background} we introduce some basic concepts and review some related work. In Section~\ref{sec:methodology} we present the overall approach. Next, in Section~\ref{sec:evaluation} we present the results of a qualitative evaluation of Pragmos. We conclude the paper providing some directions for further research in Section~\ref{sec:conclusions}.

\section{Background}
\label{sec:background}
\subsection{Large Language Models}

Language models emerged in the early 1990s as statistical models based on sequences of words~\cite{Rosenfeld00}, and later as sequences of n-grams~\cite{BrownPdLM92}, enabling already robust methods to generate sentences in natural language. However, the boom of this kind of models came with the emergence of the so-called attention-based deep neural networks, a.k.a. transformers~\cite{VaswaniSPUJGKP17}. Probably, the most notable landmark is the introduction of ChatGPT to the world, in November 2022~\cite{openai2022introducing}. Trained with an enormous corpus, hence the generic name of Large Language Model (LLM), ChatGPT show the potential of this technology because of the unprecedented capabilities for text generation, summarization, translation and even some (basic) reasoning capabilities. Sure enough, ChatGPT was followed by other similar systems which include Gemini, Llama, among others.

It was soon found that the first generation of LLMs showed very limited reasoning capabilities, primarily because its main task is to predict the next token based on statistical associations rather than engaging in logical, causal, or abstract thought. However, some researchers found that providing some examples (a.k.a. few-shot prompting~\cite{fewshot}) or a plan to be followed while tackling a problem (a.k.a. chain-of-thought -CoT- prompting~\cite{CoT}) improved radically the reasoning capabilities of LLMs. The latter gave rise to what is now called prompt engineering.

A new generation of LLMs emerged in September 2024, with OpenAI-o1 series. This LLM, in addition to the conventional text generation cycle, introduced a reasoning cycle where it takes time to emit a plan which will later be followed to produce the final outcome. Put in simple words, this new generation of LLMs standardized the use the chain-of-thought (CoT) prompting as an integral, internal part of the LLM. To distinguish the approaches, the CoT used internally by reasoning models has been called long CoT, whereas the original one is referred to as short CoT~\cite{LCoTvsSCoT}. With this evolution, a debate has emerged in whether using both long and short CoT simultaneously is adequate, because it might raise issues like ``overthinking'' and ``inference-time scaling''. Undeniably, the use of reasoning models is becoming the norm.

\subsection{Process model synthesis from behavioral relations}
As stated in the introduction, our approach is based on prior work on structuring process models. More specifically, we leverage a method for the synthesis of process models using behavior relations. These Binary behavioral relations express the execution order of activities within a process model. Given two activities, $a$ and $b$, the following three types of behavioral relations are typically considered:

\begin{enumerate}[i)]
\item \emph{Causality}: This relation indicates that the execution of one activity is a prerequisite for the execution of another. For example, $a<b$ denotes that activity $a$ precedes activity $b$.
\item \emph{Conflict}: This relation implies that the execution of one activity within a process instance excludes the execution of another. For example, $a\#b$ signifies that activities $a$ and $b$ are mutually exclusive.
\item \emph{Concurrency}: This relation indicates that two activities may be executed independently of each other, potentially in parallel (viz., concurrent), or in any order. We denote the fact that activities $a$ and $b$ are concurrent with $a\!\parallel\!b$.
\end{enumerate}


The above behavior relations can be represented with a directed graph, which is called the ordering relations graph (ORG)~\cite{PolyvyanyyGD10,PolyvyanyyGD12}. Figure~\ref{fig:org} presents one example of an ordering relations graph. There, we can see that the node $a$ has an outgoing edge that incides node $b$, which corresponds to the fact that $a$ precedes $b$. Similarly, we see that $b$ precedes $e$. Since causality is a transitive relation, we can also state that $a$ precedes $e$ and, for that reason, there is also an edge from $a$ to $e$ in the figure. Concurrency is a symmetric, irreflexive relation, which is represented with a double headed (bidirectional) edge. In our example, activities $b$ and $c$ are concurrent, as shown in Figure~\ref{fig:org}. Finally, conflict is represented by the absence of edges. For instance, $d$ is \emph{in conflict} with both $b$ and $c$ in Figure~\ref{fig:org}.

\begin{figure}[h]
    \centering
    \begin{subfigure}{0.18\textwidth}
        \scalebox{0.72}{
    \begin{tikzpicture}[every node/.style={draw,circle,minimum size=4.7mm, inner sep=0}]
        \node[draw,circle] (a) {a};
        \node[draw,circle,above right=10mm and 8mm of a] (b) {b};
        \node[draw,circle,above right=1mm and 8mm of a] (c) {c};
        \node[draw,circle,below right=3mm and 8mm of a] (d) {d};
        \node[draw,circle,right=18mm of a] (e) {e};
        
        \draw[-latex] (a) edge (b) (a) edge (d)
             (c) edge (e) (d) edge (e) (a) edge (c) (a) edge (e) (b) edge (e);
        \draw[latex-latex] (b) edge (c);
    \end{tikzpicture}  
    }
    \subcaption[]{\label{fig:org}}
    \end{subfigure}
    \begin{subfigure}{0.25\textwidth}
        \scalebox{0.72}{
        \begin{tikzpicture}[every node/.style={draw,circle,minimum size=4.7mm, inner sep=0}]
            \node (a) {a};
            \node[above right=3mm and 7mm of a] (b) {b};
            \node[below right=2mm and 12mm of a] (d) {d};
            \node[right=5mm of b] (c) {c};
            \node[right=26mm of a] (e) {e};
            \node[below=0.2mm of b,minimum size=1mm,draw=none] (hidden) {};
            \node[draw=none,above right=-0.6mm and 1mm of d] {\bf \textcolor{blue}{$C_2$}};
            
            \node[dashed,rounded corners=10,rectangle, inner sep=1.mm, fit=(b) (c) (hidden),label={[yshift=-8.5mm]\bf \textcolor{blue}{$C_1$}}] (m1) {};
            \node[dashed,rounded corners=10,rectangle, inner sep=1mm, minimum height=22mm, fit=(a) (b) (c) (d) (e) (m1),label={[xshift=-12mm,yshift=-21mm]\bf \textcolor{blue}{$L_1$}}] (m3) {}; 
            
            \path[draw,rounded corners,dashed] ([yshift=-1mm]d.south) to[out=180,in=200] ([yshift=5mm,xshift=2mm]m1.west)
            to ([yshift=5mm,xshift=-2mm]m1.east)
            to[out=-20,in=-5] ([yshift=-1mm]d.south);
            
            \draw[-latex,thick] (a) edge +(6mm,0);
            \draw[latex-,thick] (e) -- +(-6mm,0);
            \draw[latex-latex] (b) edge (c);
        \end{tikzpicture}
        }
        \subcaption[]{}
    \end{subfigure}
    \begin{subfigure}{0.17\textwidth}
    \scalebox{0.72}{
        \begin{tikzpicture}[level distance=6mm,sibling distance=10mm]
            \node[draw,rectangle] {\textcolor{blue}{$L_1$}}
                child {node {a}}
                child {node[draw] {\textcolor{blue}{$C_2$}}
                    child {node[draw] {\textcolor{blue}{$C_1$}}
                        child {node {b}}
                        child {node {c}}
                    }
                    child {node {d}}}
                child {node {e}};
        \end{tikzpicture}
        }
    \subcaption[]{}
    \end{subfigure}
    \begin{subfigure}{0.36\textwidth}
    \scalebox{0.6}{
        \begin{tikzpicture}[thick,node distance=3mm,every node/.style={font=\normalsize,scale=1}]
\node [start event] (start) {};
\node [task,right=of start,minimum width=6mm] (a) {a};
\node [diamond,draw,right= 4mm of a,minimum size=5mm,inner sep=0.5] (x1) {\large\texttimes};
\node [task,below right=0mm and 6mm of x1,minimum width=6mm] (d) {d};
\node [task,above right=12mm and 6mm of x1,minimum width=6mm] (b) {b};
\node [task,above right=3mm and 6mm of x1,minimum width=6mm] (c) {c};
\node [diamond,draw,right= 16mm of x1,minimum size=5mm,inner sep=0.5] (x2) {\large\texttimes};

\node [diamond,draw,above right= 9mm and -1mm of x1,minimum size=5mm,inner sep=0.5] (a1) {\large+};
\node [diamond,draw,above left= 9mm and -1mm of x2,minimum size=5mm,inner sep=0.5] (a2) {\large+};
\node [task,right=of x2,minimum width=6mm] (e) {e};
\node [end event,right=of e] (end) {};

\node[draw,thin,dashed,rounded corners=6,rectangle, inner sep=1mm, fit=(a1) (b) (c) (a2),label={[xshift=9mm,yshift=5mm]left:{\bf \textcolor{blue}{$C_1$}}}] (m1) {};
\node[draw,thin,dashed,rounded corners=6,rectangle, inner sep=1mm, minimum height=22mm, fit=(x1) (m1) (x2) (d),label={[xshift=-9mm,yshift=-31mm]\bf \textcolor{blue}{$C_2$}}] (m2) {}; 
\node[draw,thin,dashed,rounded corners=6,rectangle, inner sep=1mm, minimum height=22mm, fit=(a) (e) (m2),label={[xshift=-21mm,yshift=-8mm]\bf \textcolor{blue}{$L_1$}}] (m3) {};

\draw[-latex,thick] (start) edge (a) (a) edge (x1) (a1) edge (b.west) (b) edge (a2) (a2) edge (x2) (x1) edge (a1) (x1) edge (d) (d) edge (x2) (x2) edge (e) (e) edge (end)
(a1) edge (c.west) (c) edge (a2)
;

\end{tikzpicture}
}
    \subcaption[]{}
    \end{subfigure}
    \caption{Caption}
    \label{fig:enter-label}
\end{figure}

\cite{PolyvyanyyGD10} and \cite{PolyvyanyyGD12} present a method for synthesizing a BPMN model from an ordering relation graph. The method proceeds by incrementally decomposing the ORG into subgraphs, which can be classified in one of the following types:
\begin{inparaenum}[i)]
    \item a \emph{trivial}, corresponding with a single node,
    \item a \emph{complete} whenever the subgraph corresponds to a complete (sub)graph (viz., a clique) or, symmetrically, a fully disconnected (sub)graph (viz., an anti-clique or an independent set),
    \item a \emph{linear} if the subgraph is a total order (viz., a transitive relation), and
    \item a \emph{primitive} otherwise. 
\end{inparaenum}
Later, each subgraph is translated into a single-entry/single-exit (SESE) BPMN fragment. Thus, a linear is translated into a sequence, a complete into either a block of concurrency or a block of conflict (viz., a block delimited with data-exclusive gateways), and a primitive into a block with arbitrary topology.

The decomposition of an ordering relation graph as described above relies on the modular decomposition of directed graphs~\cite{McConnellM05}. Let $G = (V, E)$ be an ordering relation graph. A module $M \subseteq V$ of $G$ is a non-empty subset of vertices that exhibit the same relation with all vertices in $V \setminus M$. For example, the nodes $\{b, c\}$ in Figure~\ref{fig:org} are preceded by $b$ and have $e$ as their successor (causality relation). Moreover, neither $b$ nor $c$ has an edge to/from node $d$ (conflict relation). Hence, $\{b, c\}$ forms a complete module which is called $C_1$ in Figure~\ref{fig:org}. If we replace the nodes $\{b, c\}$ by a new node $C_1$, we will clearly see that the set of nodes $\{d, C_1\}$ forms another complete module, i.e., $C_2$, because they are disconnected, and both are preceded by $a$ and succeeded by $e$. If we replace now $\{d, C_1\}$ by a new node $C_2$, we will have the sequence of nodes $a$, $C_2$ and $e$, which forms a total order and, hence, corresponds to a linear module, i.e., $L_1$. The above illustrates the essence of the modular decomposition. There exist algorithms to compute the modular decomposition tree in linear time~\cite{McConnellM05}.

The above process implicitly constructs a rooted tree, which is known as the modular decomposition tree (MDT). By traversing the tree, we can generate a BPMN model using the following algorithm~\cite{PolyvyanyyGD10}:

\begin{algorithm}
\caption{Algorithm 1: Structuring an Acyclic Rigid Process Component}
\hspace*{\algorithmicindent} \textbf{Input} ORG: An ordering relations graph\\
\hspace*{\algorithmicindent} \textbf{Output} T: An RPST consisting of trivials, bonds, polygons
\begin{algorithmic}
\Statex
\State MDT $\gets$ ModularDecomposition(ORG) 
\State T $\gets$ RPST obtained by traversing each module M of the MDT (in
postorder) and applying the following rules:
\begin{enumerate}
    \item If M is and complete, generate an and bond component in T
    \item If M is xor complete, generate an xor bond component in T
    \item If M is linear, generate a trivial or polygon component in T
    \item If M is non-concurrent primitive, generate a well-structured
component using compiler techniques, e.g.,~\cite{Oulsnam82}
    \item If M is a concurrent primite, generate a unstructured component
    using the technique described in~\cite{PolyvyanyyGFW14}
\end{enumerate}
\State {\bf return} T
\end{algorithmic}
\end{algorithm}

Algorithm 1 allows one to generate a BPMN process model from an ordering relation graph. However, the case of primitive modules is not covered. Although for some cases, generating a BPMN fragment could be simple, for instance when it only involves concurrency or conflict, the cases where a combination of both is definitely not straightforward. However, \cite{PolyvyanyyGFW14} presents a method to tackle this problems, which relies in methods for synthesis of Petri nets, subgraph isomorphism and folding of labeled Petri nets. We refer the reader to~\cite{PolyvyanyyGFW14} for further details on the method.

\subsection{Related work}

Early attempts of using natural language processing techniques for deriving process models from textual descriptions thereof have been explored by the research community of business process management for more than one decade (e.g., \cite{FriedrichMP11,Leopold13}). The arrival of the so-called foundational models and, more specifically, large language models such as ChatGPT~\cite{ChatGPT22}, marked a milestone with a huge leap in tackling this problem.

ProMoAI is a prominent example of a tool that provides not only process modeling capabilities but also support for user-guided model redesign. ProMoAI takes as input a process description and, using an LLM with a prompt, generates Python code. This code is intended to generate an intermediate representation based on POWL, a formalism that encodes process behavior that has been developed for process model discovery from event logs. In this way, ProMoAI reuses POWL-related tools to generate a Petri net or a BPMN model, by a series of calls to Python's POWL API. Depending on the inherent complexity of the process, the resulting model might initially be inaccurate. Therefore, ProMoAI may incur in several iterations using prompts to try to fix the problems until the model produced is error-free. At this point, the tool enters into a feedback loop, within which the business analyst can refine the process model, by providing instructions in natural language to guide ProMoAI in the modification of the target process model. It is worth noting that the business analyst sees only the process model produced. The details on how the model is produced from the process description are hindered behind a highly technical process, including the Python code that is used in producing POWL and the transformation of POWL into the Petri net or BPMN model. Even if the Python code were made available to business analysts, it would require very specialized knowledge to modify such code directly. Hence, the process of model generation remains opaque.

BPMN-Chatbot is, as its name hints, a chatbot that assists business analysts in deriving process models from textual description. This tool also relies on LLMs to derive the elements of a BPMN process models, which later is assembled using ad-hoc Python code. As with ProMoAI, the tool enters a loop where the business analyst can refine the initial process model. However, the tool is capable of generating BPMN in the standard XML format. This feature allows the tool to include a Javascript frontend using the bpmn-js library, which provides high-quality rendering and auto-layout capabilities. It is not clear, however, whether the tool provides guaranties of correctness of the model that is produced.

Another interesting tool is the one presented in~\cite{KlievtsovaBKMR23}. The tool evolved from a direct prompt for ChatGPT to translate a text into a model, up to the AI-assistant for proess modeling now part of SAP Signavio. In~\cite{KlievtsovaBKMR23}, the authors propose a set of prompts to cover several tasks related to process description and corresponding model. For instance, they include a prompt intended to review and, if applicable, improve the process description, another one to enumerate the modeling elements found in the description, and, among them, one prompt which instructs the LLM to generate a model. The prompts are specified as templates such that it is possible to produce process models using a different notation by including a description of the notation. It is the last prompt which is closer to our problem. The prompt is simple and works well when the target model is also simple and considers one single step to produce the target model. Hence, it might be insufficient when the target process requires advanced workflow patterns or sophisticated control flow topologies.

We contend that the process of modeling assisted by agentic AI must allow business analysts to track closely each one of the steps. However, most of the tools known to the authors, produce technical artifacts often difficult to understand. Therefore, we take another approach trying to open up the process, by formulating the prompts with intuitive concepts and decomposing the analysis in several steps. We believe that opening-up the tool could allow business analysts to be included ``on-the-loop'', making the modeling an open-ended activity, with direct feedback and control by the analysts.
\section{Approach}

We consider modeling to be a complex task that is difficult to tackle all in a single step. In contrast to other methods, we decompose this complex task into a collection of sub-tasks that are performed incrementally to build the whole process model. Each step gathers complementary information that allows us to incrementally refine, complement, and even correct the  current version of the model. The modeling task hence produces a number of intermediate versions of the model that can be presented to human experts. This approach opens up the opportunity for human experts to check the model at different stages, question the intermediate model, and even change some of the design decisions taken by the tool.

\begin{figure}
    \centering
    \includegraphics[width=0.95\linewidth]{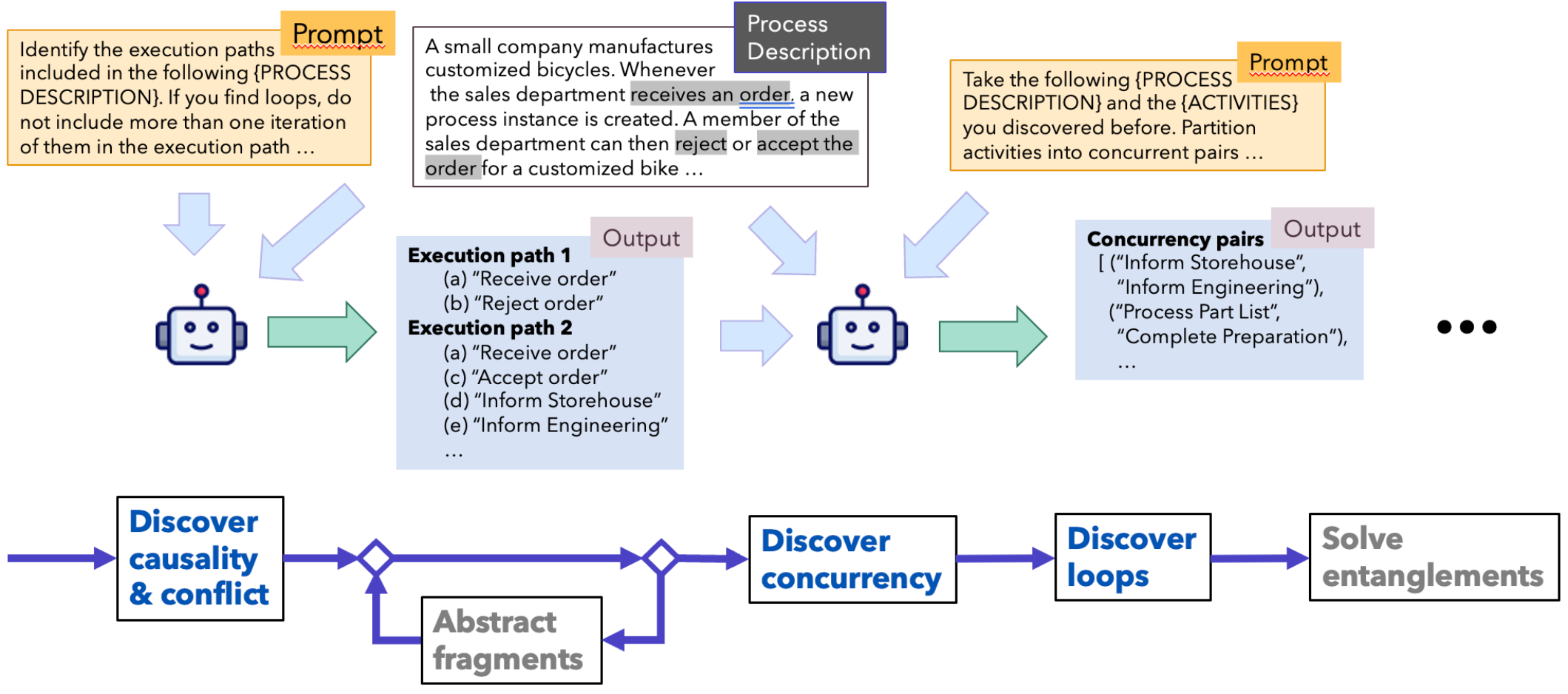}
    \caption{Pragmos' overall approach}
    \label{fig:approach}
\end{figure}

The overall workflow behind our approach is captured in Fig.~\ref{fig:approach}. On the bottom, we can see a series of boxes with the name of the steps: in blue font the elementary steps, and in gray font the steps that are needed in some special situation. In the following subsections, we describe each one of the steps.

First, the elementary steps (blue ones). We start by discovering execution paths from the process description (a bot on top of the corresponding box illustrates the input parameters and the corresponding output). These execution paths are the input for the machinery that translates them into behavioral relations, namely causality and conflict relations. It is worth noting that our system requests the LLM to avoid loops in the execution paths, as they will be processed at the end of the flow. The system generates a first model based on causality and conflict relations. This is probably the main difference with other existing systems: execution paths are intuitive enough for the analysts to understand, validate, and refine. Next, the system uncovers concurrency relations in the form of pairs of activities that can be executed simultaneously. The final steps correspond to discovering blocks of the model that have to be executed within a loop. 

In many cases, these three steps would be enough to derive a correct model from a process description. However, in other cases some additional steps would be required, the boxes in the workflow shown with a gray font. First, abstraction is often required to deal with situations where loops cannot be handled in the first step or to modularize very large descriptions. Some additional structural difficulties (entanglements) are handled in the last step: changing while loops into repeat-like loops, and uncovering optional tasks (task skipping).

\subsection{Elementary steps}

As previously shown, a BPMN model can be decomposed into a collection of single-entry/single-exit (SESE) fragments (or subgraphs). Those fragments can be one of the following: a sequence, a block of concurrency, a block of conflict, and/or a loop. The fragments could also include non-structured blocks, such as the loops with multiple entry points and/or multiple exit points, however, we will focus here only on the basic, so called structured SESE fragments.

\subsubsection{Causality and conflict relations}

\begin{figure}
    \centering
\scalebox{0.86}{
\input{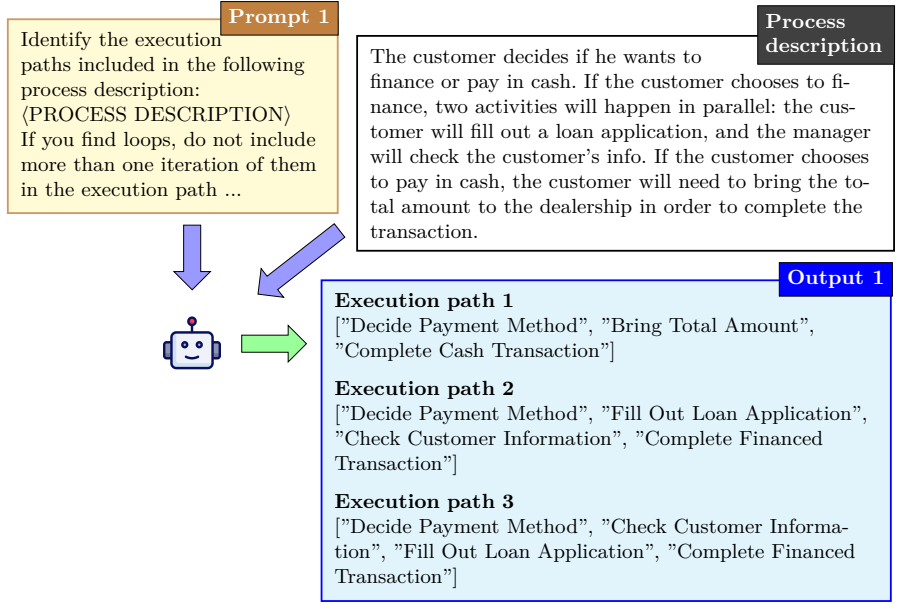}
}
\caption{\label{fig:disc:execpaths}Discovery of execution paths}
\end{figure}

As the first step, we ask the LLM to identify all possible execution paths implied in the process description. An execution path refers to a sequence of activities that are executed one after the other, a simile for execution trace as used in process mining. The idea is illustrated in Figure~\ref{fig:disc:execpaths}.

The expected result of above prompt is intuitive, as it captures the sequence of activities that will be observed during an actual execution of the process. In the example at hand, for which we have included the full process description, the LLM discovers three different execution paths. The first path, corresponds with the cases when the customer decides to pay with cash and includes the sequence: ["Decide Payment Method", "Bring Total Amount",
"Complete Cash Transaction"]. The second and third paths are followed when the customer decides to finance the payment. Note that both paths are quite similar, differing only in the order in which activities \{"Fill Out Loan Application",
"Check Customer Information"\} are executed. That is commonly interpreted as both activities being concurrent.

\begin{figure}
    \centering
\begin{tikzpicture}
\node at (0,0) (act_labels) {
\begin{tabular}{cl}
\toprule
  & Activities \\
\midrule
a & Decide Payment Method \\
b & Bring Total Amount \\
c & Complete Cash Transaction \\
d & Fill Out Loan Application \\
e & Check Customer Information \\
f & Complete Financed Transaction \\
\bottomrule
\end{tabular}
};
\node[inner sep=0pt,below=-3mm of act_labels,text width=4cm]
    {{\captionof{table}{Activity labels}}\label{tab:labels}};
\node at (6,0) (dfg) {
\begin{tikzpicture}
\node[draw,thick,minimum width=5mm,minimum height=5mm] at (0,0) (a) {a};
\node[draw,thick,minimum width=5mm,minimum height=5mm,above right=6mm and 6mm of a] (b) {b};
\node[draw,thick,minimum width=5mm,minimum height=5mm,right=6mm of b] (c) {c};
\node[draw,thick,minimum width=5mm,minimum height=5mm,right=6mm of a] (d) {d};
\node[draw,thick,minimum width=5mm,minimum height=5mm,below right=6mm and 6mm of a] (e) {e};
\node[draw,thick,minimum width=5mm,minimum height=5mm,below right=0.3mm and 6mm of d] (f) {f};

\draw[very thick,->,blue] (a) edge (b.west) (b) edge (c);
\draw[very thick,->]
    (a) edge (e.west) (d) edge (f)
    (e) edge [bend right=20] (d);
\draw[very thick,->,brown] (a) edge (d) (e) edge (f)
    (d) edge [bend right=20] (e);

\node[draw,thick,circle,minimum width=4mm,left= 7mm of a] (start) {}; 
\node[draw,fill=green!20,isosceles triangle,isosceles triangle apex angle=65,minimum size=2mm,inner sep=0] at (start) {}; 
\node[draw,thick,circle,minimum width=4mm,right= 28mm of a] (end) {}; 
\node[draw,fill=red!20,minimum size=2mm,inner sep=0] at (end) {}; 

\draw[thick,dotted,->] (start) edge (a) (c) edge (end) (f) edge (end);
\end{tikzpicture}
};
\node[inner sep=0pt,below=0mm of dfg,text width=5cm]
    {{\captionof{figure}{Directly follows graph}}\label{fig:dfg1}};

\node at (0,-5) (org) {
\begin{tikzpicture}
\node[draw,thick,minimum width=5mm,minimum height=5mm] at (0,0) (a) {a};
\node[draw,thick,minimum width=5mm,minimum height=5mm,above right=6mm and 6mm of a] (b) {b};
\node[draw,thick,minimum width=5mm,minimum height=5mm,right=6mm of b] (c) {c};
\node[draw,thick,minimum width=5mm,minimum height=5mm,right=6mm of a] (d) {d};
\node[draw,thick,minimum width=5mm,minimum height=5mm,below right=6mm and 6mm of a] (e) {e};
\node[draw,thick,minimum width=5mm,minimum height=5mm,below right=0.3mm and 6mm of d] (f) {f};

\draw[thick, ->] (a) edge (b.west) (b) edge (c)
    (a) edge (d) (d) edge (f)
    (a) edge (e.west) (e) edge (f)
    (a) edge [bend left=4] (c)
    (a) edge [bend left=50] (f);

\draw[thick,<->,red] (d) edge (e);

\node[draw,thick,circle,minimum width=4mm,left= 7mm of a] (start) {}; 
\node[draw,fill=green!20,isosceles triangle,isosceles triangle apex angle=65,minimum size=2mm,inner sep=0] at (start) {}; 
\node[draw,thick,circle,minimum width=4mm,right= 28mm of a] (end) {}; 
\node[draw,fill=red!20,minimum size=2mm,inner sep=0] at (end) {}; 

\draw[thick,dotted,->] (start) edge (a) (c) edge (end) (f) edge (end);
\end{tikzpicture}
};
\node[inner sep=0pt,below=0mm of org,text width=5cm]
    {{\captionof{figure}{Ordering relations graph}}\label{fig:org1}};

\node at (6,-5) (mdt) {
\begin{tikzpicture}
\node[draw,thick,minimum width=5mm,minimum height=5mm] at (0,0) (a) {a};
\node[draw,thick,minimum width=5mm,minimum height=5mm,above right=3.5mm and 7mm of a] (b) {b};
\node[draw,thick,minimum width=5mm,minimum height=5mm,right=4mm of b] (c) {c};
\node[draw,thick,minimum width=5mm,minimum height=5mm,below right=-3.5mm and 7mm of a] (d) {d};
\node[draw,thick,minimum width=5mm,minimum height=5mm,below right=5.5mm and 7mm of a] (e) {e};
\node[draw,thick,minimum width=5mm,minimum height=5mm,below right=-0.8mm and 4mm of d] (f) {f};
\node[draw,rounded corners,fit=(b) (c)] (bc) {};
\node[draw,rounded corners,fit=(d) (e)] (de) {};
\node[draw,rounded corners,fit=(de) (f)] (def) {};
\node[draw,rounded corners,fit=(bc) (def)] (bcdef) {};
\node[draw,rounded corners,fit=(a) (bcdef)] (abcdef) {};

\draw[thick,->] (b) edge (c) (de) edge (f)
    (a) edge (bcdef)
    ;

\draw[thick,<->,red] (d) edge (e);

\node[draw,thick,circle,minimum width=4mm,left=6mm of abcdef] (start) {}; 
\node[draw,fill=green!20,isosceles triangle,isosceles triangle apex angle=65,minimum size=2mm,inner sep=0] at (start) {}; 
\node[draw,thick,circle,minimum width=4mm,right=6mm of abcdef] (end) {}; 
\node[draw,fill=red!20,minimum size=2mm,inner sep=0] at (end) {}; 

\draw[thick,dotted,->] (start) edge (abcdef) (abcdef) edge (end);
\end{tikzpicture}
};
\node[inner sep=0pt,below=0mm of mdt,text width=7cm]
    {{\captionof{figure}{Modular decomposition tree}}\label{fig:mdt1}};

\node at (3.5,-10) (bpmn) {
\begin{tikzpicture}[every node/.append style={font=\scriptsize}]
\node[draw,thick,rounded corners,align=center] (a) at (0,0) { Decide\\ Payment\\ Method};
\node[draw,thick,diamond,right=6mm of a] (xor1) {};
\node at (xor1) {$\Xor$};
\node[draw,thick,rounded corners,align=center,above right=7mm and 10mm of xor1] (b) { Bring Total\\ Amount};
\node[draw,thick,rounded corners,align=center,right=10mm of b] (c) { Complete Cash\\ Transaction};
\node[draw,thick,diamond,below right=6mm and 4mm of xor1] (and1) {};
\node at (and1) {$\PlusP$};
\node[draw,thick,rounded corners,align=center,above right=1mm and 5.5mm of and1] (d) { Fill Out Loan\\ Application};
\node[draw,thick,rounded corners,align=center,below right=1mm and 4mm of and1] (e) { Check Customer\\ Information};
\node[draw,thick,diamond,right=28mm of and1] (and2) {};
\node[draw,thick,rounded corners,align=center,right=4mm of and2] (f) { Complete\\ Financed\\ Transaction};
\node at (and2) {$\PlusP$};
\node[draw,thick,diamond,right=62mm of xor1] (xor2) {};
\node at (xor2) {$\Xor$};
\node[draw,thick,circle,minimum width=4mm,left=5mm of a] (start) {}; 
\node[draw,ultra thick,circle,minimum width=4mm,right=5mm of xor2] (end) {}; 

\draw[thick,->] (a) edge (xor1) (b) edge (c) (and2) edge (f)
    (start) edge (a) (xor2) edge (end);
\draw[thick,->] (xor1) |- (b);
\draw[thick,->] (xor1) |- (and1);
\draw[thick,->] (and1) |- (d);
\draw[thick,->] (and1) |- (e);
\draw[thick,->] (c) -| (xor2);
\draw[thick,->] (d) -| (and2);
\draw[thick,->] (e) -| (and2);
\draw[thick,->] (f) -| (xor2);
\end{tikzpicture}
};
\node[inner sep=0pt,below=0mm of bpmn,text width=7cm]
    {{\captionof{figure}{Resulting BPMN process model}}};
\end{tikzpicture}
\end{figure}

The list of execution paths generated by the LLM serves as the source for uncovering the causal, conflict and concurrency relations. To explain the underlying procedure, we build the so-called \emph{directly-follows graph} (DFG)~\cite{Aalst18} used in process mining. A DFG is directed graph where each node represents an activity and each edge captures the fact that the activity at the source of the edge preceeds the target activity in at least one of the execution paths. Fig.~\ref{fig:dfg1} presents the DFG induced from the execution paths in Fig.~\ref{fig:disc:execpaths}. To keep the graphs compact, we are replacing the activity labels with letters according to the mapping in Table~\ref{tab:labels}. To help tracking the execution paths, we used a different edge color: blue edges for execution path 1, brown for execution path 2 and black for execution path 3. Additonally, we added a start node (circle with with a green triangle inside) and an end node (circle with a red square inside).

The ordering relations graph for our example is shown in Fig.~\ref{fig:org1}. Note that we replaced the edges connecting nodes d and e, by a single double-headed edge, which represents a concurrency relation. This decision follows the heuristic used by the well-known $\alpha$ discovery algorithm~\cite{AalstWM04}. Different to a DFG, an ordering relations graph requires causality to be transitive. In our example, the transtive closure of the causality relation includes $a \prec c$ and $a \prec f$ in addition to the direct causality represented in the DFG. We add all the missing transtive edges by computing the transtive closure of the direct causality after removing the concurrency relation. Note that each biffucation in the graph corresponds to a point where a conflict or a concurrency relation starts.

With the ORG, we compute the modular decomposition tree shown in Fig.~\ref{fig:mdt1}. First, a (concurrent) complete module can identified on the clique formed by activities $d$ and $e$. Note that both $d$ and $e$ preceed activity $f$. That is why we have a linear module, let us call it L1, comprising the clique formed by $d$, $e$ and activity $f$. On the other hand, we have another linear module composed of activities $b$ and $c$. The latter module is in conflict with L1. Finally, there is another linear module that includes activity $a$ and following by the aforementioned conflict module.

The BPMN process model can be derived directly from the ORG, using Algorithm~\ref{algo:org2bpmn}. For our example, this single step suffices to extract a correct process model from the process description. However, this step is unsufficient with more sophisticated processes. After some experimentation, we found that we have to follow an incremental approach, such that in this case we would even remove the concurrency relation. The latter is due to the fact that in several scenarios the concurrency relation can be errouneously assumed in cases where a short loop, that is a loop involving only two activities, is present. This is the reason why, our prompt for this step instructs the LLM to avoid "reentering loops".

\subsubsection{Concurrency relation}

\begin{figure}
    \centering
\scalebox{0.86}{
\input{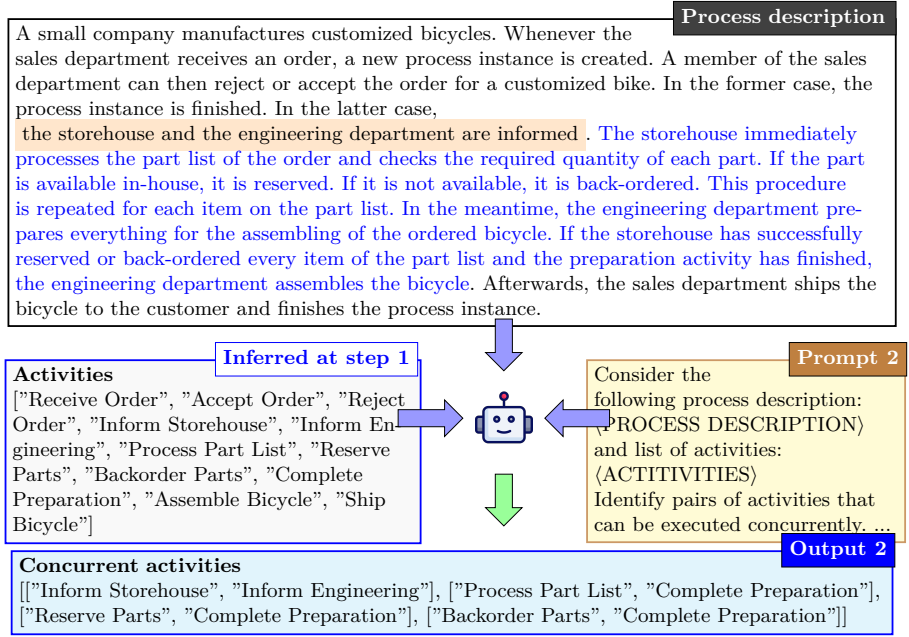}
}
\caption{\label{fig:disc:conc}Discovery of concurrency relation}
\end{figure}

In the second step of our approach, we ask the LLM to discover the concurrency relation among the activities identified in the previous step. Fig.~\ref{fig:disc:conc} summarizes the elements used in this step.

Note that in this second step, we will continue referring to the same process description. Additionally, we inject the list of activities discovered in the previous step into the new prompt. This allows us to keep the consistency in the naming of activities and, inherently, in the level of detail chosen by the LLM in the first step.

Although the previous step could reveal some concurrency, we cannot be sure that all concurrency is unveiled or even whether the revealed concurrency relation is not, in reality, induced by short loops. That is why we drop any apparent concurrency from the previous step and use the above prompt to ask the LLM to analyze the process description to discover all pairwise concurrency relations. It is worth noting that we are using a different process description from this subsection and in some subsequent ones. The underlying process model is therefore more complex and well suited for our presentation. We assume that the first step (prompt) has been performed and the resulting BPMN model is the one presented in Fig.~\ref{fig:bpmn2}, using the mapping of activities with letters according to Table~\ref{tab:labels2}.

\begin{table}
\begin{tikzpicture}
\node at (0,0) (act_labels1) {
\begin{tabular}{cl}
\toprule
  & Activities \\
\midrule
a & "Receive Order"\\
b & "Reject Order"\\
c & "Accept Order"\\
d & "Inform Storehouse"\\
\bottomrule
\end{tabular}
};

\node[right=8mm of act_labels1] (act_labels2) {
\begin{tabular}{cl}
\toprule
  & Activities \\
\midrule
e & "Inform Engineering"\\
f & "Process Part List"\\
g & "Reserve Parts"\\
h & "Backorder Parts"\\
\bottomrule
\end{tabular}
};

\node[right=8mm of act_labels2] (act_labels3) {
\begin{tabular}{cl}
\toprule
  & Activities \\
\midrule
i & "Complete Preparation"\\
j & "Assemble Bicycle"\\
k & "Ship Bicycle"\\
\\
\bottomrule
\end{tabular}
};
\end{tikzpicture}
\caption{\label{tab:labels2}Activity labels mapping}
\end{table}

\begin{figure}
\begin{tikzpicture}[every node/.style={rounded corners}]
\node[draw,circle,thick] at (0,0) (start) {};
\node[draw,thick,minimum width=5mm,minimum height=5mm,right=3mm of start] (a) {a};
\node[draw,thick,minimum width=5mm,minimum height=5mm,above right=2mm and 8mm of a] (b) {b};
\node[draw,thick,minimum width=5mm,minimum height=5mm,below right=2mm and 8mm of a] (c) {c};
\node[draw,thick,minimum width=5mm,minimum height=5mm,right=4mm of c,fill=orange!30] (d) {d};
\node[draw,thick,minimum width=5mm,minimum height=5mm,right=4mm of d,fill=orange!30] (e) {e};
\node[draw,thick,minimum width=5mm,minimum height=5mm,right=4mm of e,fill=cyan!30] (f) {f};
\node[draw,thick,minimum width=5mm,minimum height=5mm,above right=1mm and 8mm of f,fill=cyan!30] (g) {g};
\node[draw,thick,minimum width=5mm,minimum height=5mm,below right=1mm and 8mm of f,fill=cyan!30] (h) {h};
\node[draw,thick,minimum width=5mm,minimum height=5mm,right=20mm of f,fill=blue!30] (i) {i};
\node[draw,thick,minimum width=5mm,minimum height=5mm,right=4mm of i] (j) {j};
\node[draw,thick,minimum width=5mm,minimum height=5mm,right=4mm of j] (k) {k};

\node[draw,thick,diamond,rounded corners=0,right=3mm of a] (xor0) {}; 
\node at (xor0) {$\Xor$};
\node[draw,thick,diamond,rounded corners=0,right=3mm of f] (xor1) {}; 
\node at (xor1) {$\Xor$};
\node[draw,thick,diamond,rounded corners=0,right=5mm of xor1] (xor2) {}; 
\node at (xor2) {$\Xor$};
\node[draw,thick,diamond,rounded corners=0,right=78mm of xor0] (xor3) {}; 
\node at (xor3) {$\Xor$};
\node[draw,circle,ultra thick,right=3mm of xor3] (end) {};
\draw[thick,->] (xor0) |- (b);
\draw[thick,->] (xor0) |- (c);
\draw[thick,->] (xor1) |- (g);
\draw[thick,->] (xor1) |- (h);
\draw[thick,->] (g) -| (xor2);
\draw[thick,->] (h) -| (xor2);
\draw[thick,->] (b) -| (xor3);
\draw[thick,->] (k) -| (xor3);
\draw[thick,->] (start) edge (a) (a) edge (xor0) (c) edge (d) (d) edge (e) (e) edge (f)
    (f) edge (xor1) (xor2) edge (i) (i) edge (j) (j) edge (k) (xor3) edge (end);
\end{tikzpicture}
\caption{\label{tab:labels2}BPMN process model produced in step 1 from process description in Fig.~\ref{fig:disc:conc}}
\end{figure}
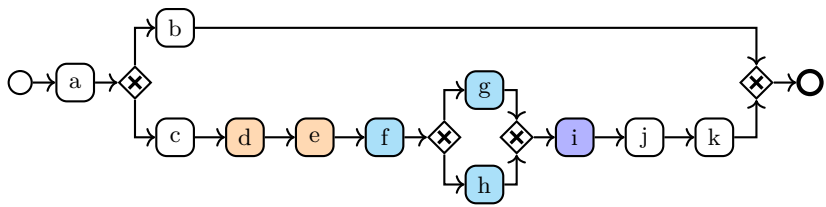

We can see the process model has already two decision points. The first one corresponds with the decision to accept (activity $c$) or reject (activity $b$) the order. In the second decision, the process selects to either reserve a part (activity $g$) or to backorder it (activity $h$). By carefully reading the process description, one can notice there are additionally two blocks with concurrency. The first one involves informing both the storehouse and the engineering department (activities $e$ and $f$ in the process model) about the order. The sentence specifying such notifications is highlighted with an orange background in the figures. The second block of concurrency is more convoluted as it involves multiple activities. However, it becomes clear when we consider that the activity "Complete preparation" (activity $i$) is made by the engineering department whereas "Process Part List", "Reserve Part" and "Backorder Part" (activities $f$, $g$ and $h$) are all done by the storehouse. Clearly, the engineering deparment can perform its activity while the storehouse is performing its own activities.

The prompt in this step asks the LLM to identify pairs of concurrent activities. The result for the example at hand is shown in Fig.~\ref{fig:disc:conc}. We have to recall that concurrency is a symmetric relation.  The LLM may enumerate all symmetrical pairs or not, however, we can complete the relation while updating the ordering relation graph.

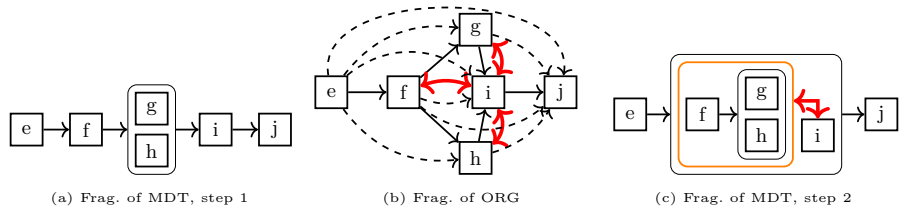
\begin{figure}
\centering
\begin{subfigure}[t]{0.32\textwidth}
\scalebox{0.85}{
\begin{tikzpicture}
    \node[draw,thick,minimum width=5mm,minimum height=5mm] at (0,0) (e) {e};
\node[draw,thick,minimum width=5mm,minimum height=5mm,right=4mm of e] (f) {f};
\node[draw,thick,minimum width=5mm,minimum height=5mm,above right=-2mm and 5mm of f] (g) {g};
\node[draw,thick,minimum width=5mm,minimum height=5mm,below right=-2mm and 5mm of f] (h) {h};
\node[draw,rounded corners,fit=(g) (h)] (gh) {};
\node[draw,thick,minimum width=5mm,minimum height=5mm,right=15mm of f] (i) {i};
\node[draw,thick,minimum width=5mm,minimum height=5mm,right=4mm of i] (j) {j};

\draw[thick,->] (e) edge (f)
    (f) edge (gh) (gh) edge (i) (i) edge (j)
    ;
\end{tikzpicture}
}
\subcaption{\label{subfig:mdt1}Frag. of MDT, step 1}
\end{subfigure}
\begin{subfigure}[t]{0.32\textwidth}
\scalebox{0.85}{
\begin{tikzpicture}
\node[draw,thick,minimum width=5mm,minimum height=5mm] at (0,0) (e) {e};
\node[draw,thick,minimum width=5mm,minimum height=5mm,right=6mm of e] (f) {f};
\node[draw,thick,minimum width=5mm,minimum height=5mm,above right=4.5mm and 6mm of f] (g) {g};
\node[draw,thick,minimum width=5mm,minimum height=5mm,below right=5mm and 6mm of f] (h) {h};
\node[draw,thick,minimum width=5mm,minimum height=5mm,right=8mm of f] (i) {i};
\node[draw,thick,minimum width=5mm,minimum height=5mm,right=6mm of i] (j) {j};

\draw[thick,->] 
    (e) edge (f) (f) edge (g)
    (f) edge (h) (g) edge (i)
    (h) edge (i) (i) edge (j)
    ;

\draw[ultra thick,<->,red] 
    (f) edge [bend left=20] (i)
    (g) edge [bend left=40] (i)
    (h) edge [bend right=40] (i)
    ;
\draw[thick,dashed,->] (g) edge [bend left=20] (j)
    (f) edge [bend right=20] (i)
    (f) edge [bend right=40] (j)
    (h) edge [bend right=20] (j)
    (e) edge [bend left=40] (i.north west)
    (e) edge [bend left=30] (g)
    (e) edge [bend left=100] (j)
    (e) edge [bend right=30] (h)
    ;
\end{tikzpicture}
}
\subcaption{\label{subfig:org}Frag. of ORG}
\end{subfigure}
\begin{subfigure}[t]{0.32\textwidth}
\scalebox{0.85}{
\begin{tikzpicture}
\node[draw,thick,minimum width=5mm,minimum height=5mm] at (0,0) (e) {e};
\node[draw,thick,minimum width=5mm,minimum height=5mm,right=6mm of e] (f) {f};
\node[draw,thick,minimum width=5mm,minimum height=5mm,above right=-2mm and 4mm of f] (g) {g};
\node[draw,thick,minimum width=5mm,minimum height=5mm,below right=-2mm and 4mm of f] (h) {h};
\node[draw,rounded corners,fit=(g) (h)] (gh) {};
\node[draw=orange,thick,rounded corners,fit=(f) (gh)] (fgh) {};
\node[draw,thick,minimum width=5mm,minimum height=5mm,right=3.5mm of h] (i) {i};
\node[draw,rounded corners,fit=(i) (fgh)] (fghi) {};
\node[draw,thick,minimum width=5mm,minimum height=5mm,right=34mm of e] (j) {j};

\draw[thick,->] 
    (e) edge (fghi) 
    (f) edge (gh)
    (fghi) edge (j)
    ;

\draw[ultra thick,<->,red] 
    ([yshift=(3mm)]fgh) -| (i)
    ;
\end{tikzpicture}
}
\subcaption{\label{subfig:mdt2}Frag. of MDT, step 2}
\end{subfigure}
\caption{\label{fig:conc:org}Frags. of modular decomposition trees and ordering relations graph of the running example}
\end{figure}

\begin{figure}
    \begin{tikzpicture}[every node/.style={rounded corners}]
    \node[draw,circle,thick] at (0,0) (start) {};
\node[draw,thick,minimum width=5mm,minimum height=5mm,right=4mm of start] (a) {a};
\node[draw,thick,minimum width=5mm,minimum height=5mm,above right=2mm and 10mm of a] (b) {b};
\node[draw,thick,minimum width=5mm,minimum height=5mm,below right=2mm and 10mm of a] (c) {c};
\node[draw,thick,diamond,rounded corners=0,right=4mm of c] (and0) {};
\node at (and0) {$\PlusP$}; 
\node[draw,thick,diamond,rounded corners=0,right=7mm of and0] (and1) {};
\node at (and1) {$\PlusP$}; 
\node[draw,thick,diamond,rounded corners=0,right=4mm of and1] (and2) {};
\node at (and2) {$\PlusP$}; 
\node[draw,thick,diamond,rounded corners=0,right=38mm of and1] (and3) {};
\node at (and3) {$\PlusP$};
\node[draw,thick,minimum width=5mm,minimum height=5mm,above right=-1mm and 10mm of c] (d) {d};
\node[draw,thick,minimum width=5mm,minimum height=5mm,below right=-1mm and 10mm of c] (e) {e};

\node[draw,thick,minimum width=5mm,minimum height=5mm,above right=2mm and 2mm of and2] (f) {f};
\node[draw,thick,minimum width=5mm,minimum height=5mm,above right=-1mm and 10mm of f] (g) {g};
\node[draw,thick,minimum width=5mm,minimum height=5mm,below right=-1mm and 10mm of f] (h) {h};
\node[draw,thick,minimum width=5mm,minimum height=5mm,below right=2mm and 2mm of and2] (i) {i};
\node[draw,thick,minimum width=5mm,minimum height=5mm,right=4mm of and3] (j) {j};
\node[draw,thick,minimum width=5mm,minimum height=5mm,right=6mm of j] (k) {k};
\node[draw,thick,diamond,rounded corners=0,right=93mm of xor0] (xor3) {}; 
\node at (xor3) {$\Xor$};
\node[draw,circle,ultra thick,right=3mm of xor3] (end) {};

\node[draw,thick,diamond,rounded corners=0,right=4mm of a] (xor0) {};
\node at (xor0) {$\Xor$};
\node[draw,thick,diamond,rounded corners=0,right=4mm of f] (xor1) {}; 
\node at (xor1) {$\Xor$};
\node[draw,thick,diamond,rounded corners=0,right=7mm of xor1] (xor2) {}; 
\node at (xor2) {$\Xor$};

\draw[thick,->] (xor0) |- (b);
\draw[thick,->] (xor0) |- (c);
\draw[thick,->] (and0) |- (d);
\draw[thick,->] (and0) |- (e);
\draw[thick,->] (and2) |- (f);
\draw[thick,->] (and2) |- (i);
\draw[thick,->] (xor1) |- (g);
\draw[thick,->] (xor1) |- (h);
\draw[thick,->] (d) -| (and1);
\draw[thick,->] (i) -| (and3);
\draw[thick,->] (xor2) -| (and3);
\draw[thick,->] (e) -| (and1);
\draw[thick,->] (g) -| (xor2);
\draw[thick,->] (h) -| (xor2);
\draw[thick,->] (b) -| (xor3);
\draw[thick,->] (k) -| (xor3);
\draw[thick,->] (start) edge (a) (a) edge (xor0) (c) edge (and0) (and1) edge (and2)
    (f) edge (xor1) (and3) edge (j) (j) edge (k) (xor3) edge (end);
\end{tikzpicture}
    \caption{\label{fig:conc:bpmn}BPMN model produced in step 2}
\end{figure}
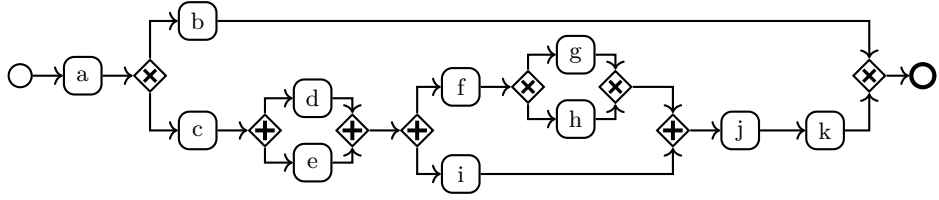

To better illustrate how concurrency is introduced, let us focus in a fragment of the process model. The corresponding ordering relations graph is shown in Fig.~\ref{subfig:org}. In the first step, causality (black, solid and dashed arrows) and conflict  (lack of edge connections) relations were discovered. The MDT induced from the ORG in the first step is shown in Fig.~\ref{subfig:mdt1}. The concurrency discovered in this second step is presented with red, double headed arrows in the ORG in Fig.~\ref{subfig:org}. Therefore, the idea is to replace the causal relation between each pair of activities by a concurrency relation, i.e., $f \prec i$ is replaced by $f \parallel i$. Once all the concurrency relations are replaced, a new MDT can be computed, as shown in Fig.~\ref{subfig:mdt2}. It can be easily verified in the new MDT that activity $i$ is concurrent with a module that contains all the activities $f$, $g$ and $h$ as expected, while preserving the causality observed between activity $f$ and the block of conflict already in place between activities $g$ and $h$. Applying this procedure to entire ORG, we will be able to derive the BPMN model shown in Fig.~\ref{fig:conc:bpmn}.

After our trials, we could see that cases leading to very simple blocks of concurrency, like the one comprising activities $d$ and $e$ could be discovered from the set of execution paths generated in step 1. However, situations like the block of concurrency involving activities $f$, $g$, $h$ and $i$, which additionally include conflict, are difficult to be solved in step 1. Separating the discovery of the behavior relations separately seems to be appropriate. Moreover, introducing concurrency in the ordering relations graph is straightforward, and generating the BPMN out of the ORG is properly handled by Algorithm~\ref{algo:org2bpmn}.

\subsubsection{Discovering Loops}

\begin{figure}
    \centering
\scalebox{0.86}{
\input{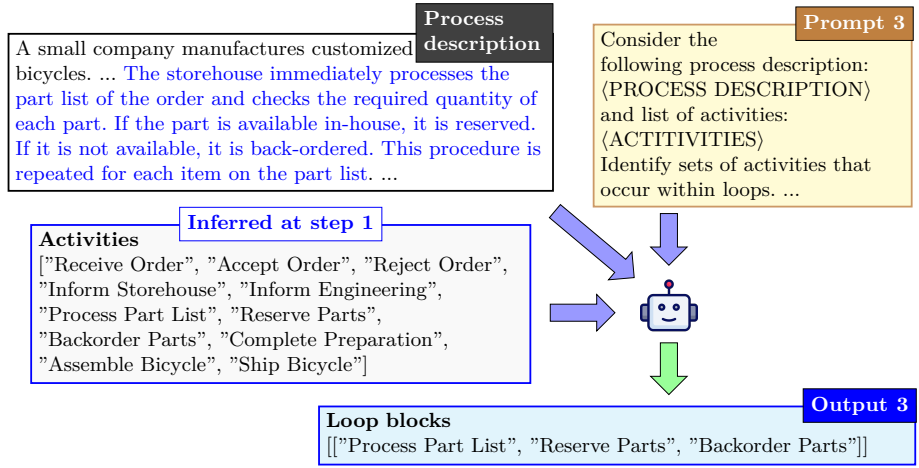}
}
\caption{\label{fig:disc:loops}Discovery of loops}
\end{figure}

The third step in the procedure corresponds with discovering fragments of the process incurring in repetitive behavior. As stated before, the presence of loops may induce some difficulties in the discovery of the process model. Indeed, the presence of short loops can be confounded with concurrency. That is the main reason we ask the LLM to avoid reentering loops in the first step. Hence, in this third step we ask the LLM to identify blocks (groups of activities) that make part of a loop. The elements used in this step are shown in Fig.~\ref{fig:disc:loops}.

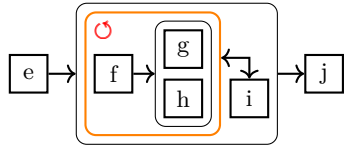
\begin{figure}
    \centering
    \begin{tikzpicture}
\node[draw,thick,minimum width=5mm,minimum height=5mm] at (0,0) (e) {e};
\node[draw,thick,minimum width=5mm,minimum height=5mm,right=6mm of e] (f) {f};
\node[draw,thick,minimum width=5mm,minimum height=5mm,above right=-2mm and 4mm of f] (g) {g};
\node[draw,thick,minimum width=5mm,minimum height=5mm,below right=-2mm and 4mm of f] (h) {h};
\node[draw,rounded corners,fit=(g) (h)] (gh) {};
\node[draw=orange,thick,rounded corners,fit=(f) (gh)] (fgh) {};
\node[draw,thick,minimum width=5mm,minimum height=5mm,right=3.5mm of h] (i) {i};
\node[draw,rounded corners,fit=(i) (fgh)] (fghi) {};
\node at ([shift={(2.5mm,-2.5mm)}]fgh.north west) {\textcolor{red}{$\bm{\circlearrowleft}$}};
\node[draw,thick,minimum width=5mm,minimum height=5mm,right=34mm of e] (j) {j};

\draw[thick,->] 
    (e) edge (fghi) 
    (f) edge (gh)
    (fghi) edge (j)
    ;

\draw[thick,<->] 
    ([yshift=(3mm)]fgh) -| (i)
    ;
\end{tikzpicture}
    \caption{\label{fig:loop:mdt1}MDT annotated with a loop}
\end{figure}

It is worth recalling that an ordering relations graph and its modular decomposition can only be associated to an acyclic process model. To overcome this limitation, here we annotate modules in a modular decomposition tree to indicate that the underlying activities are enclosed within a repeat-like loop. Let us consider the running example and the output of the third prompt, shown in Fig.~\ref{fig:disc:loops}. We can see that the process description contains a single loop, involving activities "Process Part List", "Reserve Parts", "Backorder Parts" (i.e. activities $f$, $g$ and $h$). Therefore, we annotate the MDT as shown in Fig.~\ref{fig:loop:mdt1}. To help the reader identify the module that was annotated with a loop, the graphical representation is colored in the figure.

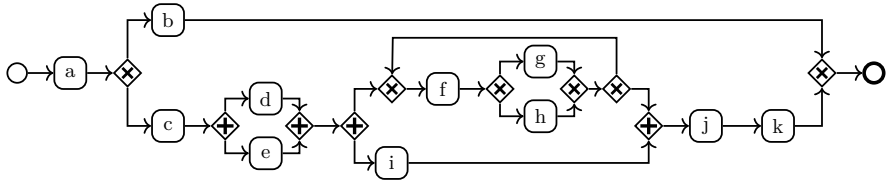
\begin{figure}
    \centering
    \scalebox{0.85} {
    \begin{tikzpicture}[every node/.style={rounded corners}]
\node[draw,circle,thick] at (0,0) (start) {};
\node[draw,thick,minimum width=5mm,minimum height=5mm,right=4mm of start] (a) {a};
\node[draw,thick,minimum width=5mm,minimum height=5mm,above right=3mm and 10mm of a] (b) {b};
\node[draw,thick,minimum width=5mm,minimum height=5mm,below right=3mm and 10mm of a] (c) {c};
\node[draw,thick,diamond,rounded corners=0,right=4mm of c] (and0) {};
\node at (and0) {$\PlusP$}; 
\node[draw,thick,diamond,rounded corners=0,right=7mm of and0] (and1) {};
\node at (and1) {$\PlusP$}; 
\node[draw,thick,diamond,rounded corners=0,right=4mm of and1] (and2) {};
\node at (and2) {$\PlusP$}; 
\node[draw,thick,diamond,rounded corners=0,right=5cm of and1] (and3) {};
\node at (and3) {$\PlusP$};
\node[draw,thick,minimum width=5mm,minimum height=5mm,above right=-1mm and 10mm of c] (d) {d};
\node[draw,thick,minimum width=5mm,minimum height=5mm,below right=-1mm and 10mm of c] (e) {e};

\node[draw,thick,minimum width=5mm,minimum height=5mm,above right=2mm and 10mm of and2] (f) {f};
\node[draw,thick,minimum width=5mm,minimum height=5mm,above right=-1mm and 10mm of f] (g) {g};
\node[draw,thick,minimum width=5mm,minimum height=5mm,below right=-1mm and 10mm of f] (h) {h};
\node[draw,thick,minimum width=5mm,minimum height=5mm,below right=2mm and 2mm of and2] (i) {i};
\node[draw,thick,minimum width=5mm,minimum height=5mm,right=4mm of and3] (j) {j};
\node[draw,thick,minimum width=5mm,minimum height=5mm,right=6mm of j] (k) {k};

\node[draw,thick,diamond,rounded corners=0,right=4mm of a] (xor0) {};
\node at (xor0) {$\Xor$};
\node[draw,thick,diamond,rounded corners=0,right=4mm of f] (xor1) {}; 
\node at (xor1) {$\Xor$};
\node[draw,thick,diamond,rounded corners=0,right=7mm of xor1] (xor2) {}; 
\node at (xor2) {$\Xor$};
\node[draw,thick,diamond,rounded corners=0,left=3mm of f] (rep1) {}; 
\node at (rep1) {$\Xor$};
\node[draw,thick,diamond,rounded corners=0,right=2mm of xor2] (rep2) {}; 
\node at (rep2) {$\Xor$};
\node[draw,thick,diamond,rounded corners=0,right=104mm of xor0] (xor3) {}; 
\node at (xor3) {$\Xor$};
\node[draw,circle,ultra thick,right=4mm of xor3] (end) {};

\draw[thick,->] (xor0) |- (b);
\draw[thick,->] (xor0) |- (c);
\draw[thick,->] (and0) |- (d);
\draw[thick,->] (and0) |- (e);
\draw[thick,->] (and2) |- (rep1);
\draw[thick,->] (and2) |- (i);
\draw[thick,->] (xor1) |- (g);
\draw[thick,->] (xor1) |- (h);
\draw[thick,->] (d) -| (and1);
\draw[thick,->] (i) -| (and3);
\draw[thick,->] (rep2) -| (and3);
\draw[thick,->] (e) -| (and1);
\draw[thick,->] (g) -| (xor2);
\draw[thick,->] (h) -| (xor2);
\draw[thick,->] (b) -| (xor3);
\draw[thick,->] (k) -| (xor3);
\draw[thick,->] (a) edge (xor0) (c) edge (and0) (and1) edge (and2)
    (rep1) edge (f) (xor2) edge (rep2)
    (f) edge (xor1) (and3) edge (j) (j) edge (k)
    (start) edge (a) (xor3) edge (end)
    (rep2) -- ++(0,8mm) -- ++(-35mm,0) -- (rep1);
\end{tikzpicture}
    }
    \caption{\label{fig:loop:bpmn1}Final BPMN process model}
\end{figure}

With this recently added information, we can rewrite the BPMN model including the loop as shown in Fig.~\ref{fig:loop:bpmn1}. Note that we assume that the loop is always associated with a module. For example, it is not possible to have activities $f$ and $g$ in a loop, while $i$ is not part of it. This is a common structural requirement for well-formedness and can be enforced by iterating with the LLM.

It is also worth noting that the loop introduced in the BPMN model is structurally equivalent to a repeat loop in programming languages: the loop body is inconditionally executed once and some additional times until the condition at the exit gateway holds true. In this example, re-entering the loop would imply re-executing the loop body.

\subsection{Block abstraction}

In some cases, loops appear irremediably in the model because they are somehow enforced in the process description. In those cases, we use a sort of block abstraction as a way to eliminate the loops. To illustrate the situation, let us consider the following process description:

\begin{figure}
\begin{tikzpicture}
\node[draw,thick,text width=12cm] at (0,0) (procdesc) {
A customer brings in a defective computer and the CRS checks the\\
defect and hands out a repair cost calculation back. If the customer decides that the costs are acceptable, the process continues, otherwise she takes her computer home unrepaired. The ongoing repair consists of two activities, which are executed, in an arbitrary order. The first activity is to check and repair the hardware, whereas the second activity checks and configures the software. After each of these activities, the proper system functionality is tested. If an error is detected another arbitrary repair activity is executed, otherwise the repair is finished.
};
\node[draw,inner sep=1.2mm,text width=18mm,fill=darkgray] at ([shift={(-10.5mm,0)}]procdesc.north east) (pdlabel) {\textcolor{white}{\bf\small Process\\\small description}};
\end{tikzpicture}
\caption{\label{fig:loops2:abstraction}Process description: Computer repair}
\end{figure}

The set of execution paths discovered by step 1 is shown in Fig.~\ref{}. Note that the LLM identifies two activities to deal with hardware issues, i.e., ``Check hardware'' and ``Repair hardware'', and two other activities to deal with software related problems, i.e., ``Check software'' and ``Configure software''. There is another activity, i.e. ``Test system functionality'', which is performed after any hardware or software interventions. As a consequence, the LLM adds ``Test system functionality'' twice in an execution path, which will result in a loop in the discovered process model. At this point, we have considered rewriting the prompt to instruct the LLM to select a different name, e.g., ``Test software ...'' and ``Test hardware ...'', for each activity, respectively, in cases similar to this one. However, we noticed that some LLMs do not comply with this instruction all the time. That is the reason why, we decided to rely on block abstraction, which can be useful in other contexts similar to this one.

The process description we are analyzing has another particularity. The fact that a computer can be checked for hardware and software issues in any arbitrary order introduces another type of cycle. This kind of causal independence in contexts can be associated with concurrency, e.g., shipping a product while handling the invoicing, but in our example, hardware and software checking cannot be done simultaneously, but one after the other. Therefore, we need a generic approach for dealing with these two situations.

\begin{figure}
\begin{tikzpicture}
\node[anchor=north] at (0,0) (labels) {
    \begin{tabular}{p{42mm}l}
    \toprule
    Execution path 1\\
    \midrule
    Receive defective computer & a\\
    Assess computer defect & b\\
    Provide cost calculation & c\\
    Return computer unrepaired & d\\
    \bottomrule
    \end{tabular}
    };
\node[anchor=north] at (5,0) (labels) {
    \begin{tabular}{p{42mm}l}
    \toprule
    Execution path 2\\
    \midrule
    Receive defective computer & a\\
    Assess computer defect & b\\
    Provide cost calculation & c\\
    \textcolor{blue}{Check hardware} & d \\
    \textcolor{blue}{Repair hardware} & e \\
    \textcolor{blue}{Test system functionality} & f \\
    \textcolor{red}{Check software} & g \\
    \textcolor{red}{Configure software} & h \\
    \textcolor{red}{Test system functionality} & f \\
    \bottomrule
    \end{tabular}
    };
\node[anchor=north] at (10,0) (labels) {
    \begin{tabular}{p{42mm}l}
    \toprule
    Execution path 3\\
    \midrule
    Receive defective computer & a\\
    Assess computer defect & b\\
    Provide cost calculation & c\\
    \textcolor{red}{Check software} & g \\
    \textcolor{red}{Configure software} & h \\
    \textcolor{red}{Test system functionality} & f \\
    \textcolor{blue}{Check hardware} & d \\
    \textcolor{blue}{Repair hardware} & e \\
    \textcolor{blue}{Test system functionality} & f \\
    \bottomrule
    \end{tabular}
};

\node at (0,-8) (labelsp) {
    \begin{tabular}{p{42mm}l}
    \toprule
    Execution path 1\\
    \midrule
    Receive defective computer & a\\
    Assess computer defect & b\\
    Provide cost calculation & c\\
    Return computer unrepaired & d\\
    \midrule
    Execution path 2\\
    \midrule
    Receive defective computer & a\\
    Assess computer defect & b\\
    Provide cost calculation & c\\
    \textcolor{blue}{Repair hardware defect} & i \\
    \textcolor{red}{Fix software configuration} & j \\
    \midrule
    Execution path 2\\
    \midrule
    Receive defective computer & a\\
    Assess computer defect & b\\
    Provide cost calculation & c\\
    \textcolor{red}{Fix software configuration} & j \\
    \textcolor{blue}{Repair hardware defect} & i \\
    \bottomrule
    \end{tabular}
};
\end{tikzpicture}
\end{figure}

\subsection{Structural entanglements}

Consider now the process description shown in the Fig.~\ref{fig:loops2:procdesc}. One possible BPMN model for this description, manually generated by one of the authors, is shown in Fig~\ref{fig:loops2:bpmn:manual}.

\begin{figure}
\begin{tikzpicture}
\node[draw,thick,text width=12cm] at (0,0) (procdesc) {
The process begins when the student logs in to the university's web-\\site.
He then takes an online exam. After that, the system grades it. If the student scores below 60\%, he takes the exam again. If the student scores 60\% or higher on the exam, the professor enters the grade.
};
\node[draw,inner sep=1.2mm,text width=18mm,fill=darkgray] at ([shift={(-10.5mm,0)}]procdesc.north east) (pdlabel) {\textcolor{white}{\bf\small Process\\\small description}};
\end{tikzpicture}
\caption{\label{fig:loops2:procdesc}Process description: Online exam}
\end{figure}

\begin{figure}
    \centering
    \scalebox{0.85}{
\begin{tikzpicture}[every node/.style={rounded corners,align=center,font={\small}}]
\node[draw,circle,minimum width=3mm,thick] at (0,0) (start) {};
\node[draw,thick,minimum width=5mm,minimum height=5mm,right=3mm of start] (a) {Log into\\universty\\website};
\node[draw,thick,diamond,rounded corners=0,right=3mm of a] (xor0) {}; 
\node at (xor0) {$\Xor$};
\node[draw,thick,minimum width=5mm,minimum height=5mm,above right=3mm and 3mm of xor0] (b) {Complete\\online exam};
\node[draw,thick,minimum width=5mm,minimum height=5mm,right=4mm of b] (c) {Grade\\exam};
\node[draw,thick,diamond,rounded corners=0,below right=3mm and 3mm of c] (xor1) {}; 
\node at (xor1) {$\Xor$};
\node[draw,thick,minimum width=5mm,minimum height=5mm,right=4mm of xor1] (d) {Register\\grade};
\node[draw,circle,ultra thick,right=3mm of d] (end) {};
\draw[thick,->] (start) edge (a) (a) edge (xor0) (b) edge (c)
    (xor1) edge (d) (d) edge (end);
\draw[thick,->] (xor0) |- (b);
\draw[thick,->] (c) -| (xor1);
\draw[thick,->] (xor1.south) -| ++(0,-5mm) -- ([yshift=-5mm]xor0.south) -- (xor0.south);
\end{tikzpicture}
}
\caption{\label{fig:loops2:bpmn:manual} BPMN model, manually derived from description in Fig.~\ref{fig:loops2:procdesc}}
\end{figure}
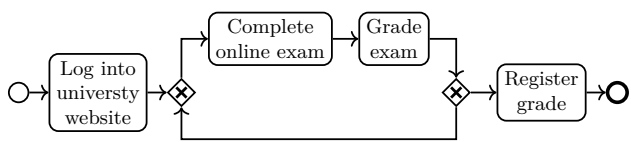

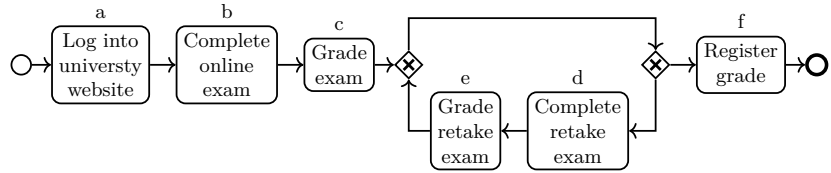
\begin{figure}
    \centering
\scalebox{0.85}{
\begin{tikzpicture}[every node/.style={rounded corners,align=center,font={\small}}]
\node[draw,circle,minimum width=3mm,thick] at (0,0) (start) {};
\node[draw,thick,minimum width=5mm,minimum height=5mm,right=3mm of start,label=a] (a) {Log into\\universty\\website};
\node[draw,thick,minimum width=5mm,minimum height=5mm,right=4mm of a,label=b] (b) {Complete\\online\\exam};
\node[draw,thick,minimum width=5mm,minimum height=5mm,right=4mm of b,label=c] (c) {Grade\\exam};
\node[draw,thick,diamond,rounded corners=0,right=3mm of c] (xor0) {}; 
\node at (xor0) {$\Xor$};
\node[draw,thick,diamond,rounded corners=0,right=34mm of xor0] (xor1) {}; 
\node at (xor1) {$\Xor$};
\node[draw,thick,minimum width=5mm,minimum height=5mm,below left=3mm and 3mm of xor1,label=d] (d) {Complete\\retake\\exam};
\node[draw,thick,minimum width=5mm,minimum height=5mm,left=4mm of d,label=e] (e) {Grade\\retake\\exam};
\node[draw,thick,minimum width=5mm,minimum height=5mm,right=4mm of xor1,label=f] (f) {Register\\grade};
\node[draw,circle,ultra thick,right=3mm of f] (end) {};
\draw[thick,->] (start) edge (a) (a) edge (b) (b) edge (c)
    (c) edge (xor0) (d) edge (e) (xor1) edge (f) (f) edge (end);
\draw[thick,->] (xor1) |- (d);
\draw[thick,->] (e) -| (xor0);
\draw[thick,->] (xor0.north) |- ++(0,5mm) -- ([yshift=5mm]xor1.north) -- (xor1.north);
\end{tikzpicture}
}
\caption{\label{fig:loops2:bpmn_final}BPMN model generated by Pragmos}
\end{figure}

Due to some of the constraints imposed by our approach, Pragmos would generate a different, yet equally valid BPMN model: the one shown in Fig.~\ref{fig:loops2:bpmn_final}. Indeed, the prompt used in the first step instructs the LLM to avoid entering the body of any loop implied by the process description. To overcome this limitation and, at the same time, to be thorough in the analysis of the process description, the LLM relies in choosing a different label for two possibly equivalent activities. Specifically, the LLM ended up using "Complete online exam" and "Complete retake exam", which can be considered to be two instances of the same activity, and similarly "Grade exam" and "Grade retake exam". So, relabeling is a resort the LLM uses to avoid loops. However, this decision challenges Pragmos' capabilities so far. We use this example to make evident two "control flow topologies" which could emerge in other contexts, posing challenges to Pragmos, and present some extensions to the procedure so far to cope with such challenges.

\begin{figure}
    \centering
\scalebox{0.85} {
\begin{tikzpicture}
\node at (0,0) (labels) {
    \begin{tabular}{p{42mm}l}
        \toprule
        Execution path 1\\
        \midrule
        Log into university website & a\\
        Complete online exam & b\\
        Grade exam & c\\
        Register grade & f\\
        \midrule
        Execution path 2\\
        \midrule
        Log into university website & a\\
        Complete online exam & b\\
        Grade exam & c\\
        Complete retake exam & d\\
        Grade retake exam & e\\   
        Register grade & f\\
        \bottomrule
    \end{tabular}
};
\node at ([shift={(44mm,0)}]labels.north east) (dfg) {
\begin{tikzpicture}
\node[draw,thick,minimum width=5mm,minimum height=5mm] at (0,0) (a) {a};
\node[draw,thick,minimum width=5mm,minimum height=5mm,right=5mm of a] (b) {b};
\node[draw,thick,minimum width=5mm,minimum height=5mm,right=5mm of b] (c) {c};
\node[draw,thick,minimum width=5mm,minimum height=5mm,above right=0.5mm and 5mm of c] (d) {d};
\node[draw,thick,minimum width=5mm,minimum height=5mm,right=5mm of d] (e) {e};
\node[draw,thick,minimum width=5mm,minimum height=5mm,below right=0.5mm and 5mm of e] (f) {f};
\node[draw,thick,circle,minimum width=4mm,left=4mm of a] (start) {}; 
\node[draw,fill=green!20,isosceles triangle,isosceles triangle apex angle=65,minimum size=2mm,inner sep=0] at (start) {}; 
\node[draw,thick,circle,minimum width=4mm,right=4mm of f] (end) {}; 
\node[draw,fill=red!20,minimum size=2mm,inner sep=0] at (end) {};
\draw[thick,->] (start) edge (a) (a) edge (b) (b) edge (c)
    (c) edge (d.west) (d) edge (e) (e) edge (f) (f) edge (end);
\draw[thick,->,blue] (c) edge [bend right=10] (f);
\end{tikzpicture}
};

\node[below= 1cm of dfg] (bpmn) {
\begin{tikzpicture}[rounded corners]
\node[draw,thick,minimum width=5mm,minimum height=5mm] at (0,0) (a) {a};
\node[draw,thick,minimum width=5mm,minimum height=5mm,right=5mm of a] (b) {b};
\node[draw,thick,minimum width=5mm,minimum height=5mm,right=5mm of b] (c) {c};
\node[draw,thick,minimum width=5mm,minimum height=5mm,right=2mm and 5mm of c] (d) {d};
\node[draw,thick,minimum width=5mm,minimum height=5mm,right=5mm of d] (e) {e};
\node[draw,thick,minimum width=5mm,minimum height=5mm,right=2mm and 5mm of e] (f) {f};
\node[draw,thick,circle,minimum width=4mm,left=4mm of a] (start) {}; 
\node[draw,ultra thick,circle,minimum width=4mm,right=4mm of f] (end) {}; 

\draw[thick,->] (start) edge (a) (a) edge (b) (b) edge (c)
    (c) edge (d) (d) edge (e) (e) edge (f) (f) edge (end);
\end{tikzpicture}
};

\node[below= 1cm of bpmn] (bpmn2) {
\begin{tikzpicture}[rounded corners]
\node[draw,thick,minimum width=5mm,minimum height=5mm] at (0,0) (a) {a};
\node[draw,thick,minimum width=5mm,minimum height=5mm,right=4mm of a] (b) {b};
\node[draw,thick,minimum width=5mm,minimum height=5mm,right=4mm of b] (c) {c};
\node[draw,thick,minimum width=5mm,minimum height=5mm,above right=2mm and 8mm of c] (d) {d};
\node[draw,thick,minimum width=5mm,minimum height=5mm,right=4mm of d] (e) {e};
\node[draw,thick,minimum width=5mm,minimum height=5mm,below right=2mm and 7mm of e] (f) {f};

\node[draw,thick,diamond,right=3mm of c,rounded corners=0] (xor1) {};
\node at ([shift={(0,2.2mm)}]xor1) {$\Xor$};
\node[draw,thick,diamond,left=3mm of f,rounded corners=0] (xor2) {};
\node at ([shift={(0,2.2mm)}]xor2) {$\Xor$};

\node[draw,thick,circle,minimum width=4mm,left=4mm of a] (start) {}; 
\node[draw,ultra thick,circle,minimum width=4mm,right=4mm of f] (end) {}; 
\draw[thick,->] (start) edge (a) (a) edge (b) (b) edge (c)
    (c) edge (xor1) (xor2) edge (f)
    (d) edge (e) (f) edge (end);
\draw[thick,->,rounded corners=0] (xor1) |- (d);
\draw[thick,->,rounded corners=0] (e) -| (xor2);
\draw[thick,->,rounded corners=0] (xor2.south) -- ([shift=({0,-4mm})]xor2.south) -- ([shift=({0,-4mm})]xor1.south) -- (xor1.south);
\end{tikzpicture}
};

\node[inner sep=0pt,below=-1mm of dfg,text width=7cm]
    {{\caption{Directly-follows graph}}};
\node[inner sep=0pt,below=-1mm of bpmn,text width=7cm]
    {{\captionof{figure}{BPMN model, step 1}}};
\node[inner sep=0pt,below=-1mm of bpmn2,text width=7cm]
    {{\captionof{figure}{BPMN model, step 3}}};
\end{tikzpicture}
}
\caption{}
\end{figure}

Let us analyze the aforementioned challenges. If we follow the current procedure, the first step would identify two traces as shown in Fig.~\ref{}. Such paths give rise to the directly-follows graph (DFG) shown in Fig.~\ref{}. Interestingly, the DFG seems similar to the expected result. However, the directly-follows relation between activities $c$ and $f$ happens to coincide with a  transitive causality relation, which will vanish when generating the BPMN model (and other intermediate graphs), as shown in Fig.~\ref{}. The causality relations that are vanished are associated with paths that make optional one or several activities. This problem is also observed in process mining, in discovery algorithms such as the well-known $\alpha$ algorithm.

Later in step 3, the LLM properly identifies activities $d$ and $e$ as being part of a loop, allowing Pragmos to generate the BPMN model shown in Fig.~\ref{}. However, in this example a while-like loop is required and Pragmos generated a repeat-like loop. This means that activities "Complete retake exam" and "Grade retake exam" are executed at least once, which is not consistent with the process description. In fact, this problem can be fixed by adding the causality relation that was vanished, as mentioned before. However, two problems: missing causality relation and erroneous loop type, can be nicely fixed by transforming the loop into a while-loop. Note that the vanishing of causality relations can occur in other contexts, not necessarily associated with a loop. 

To cope with these two problems, we use a procedure similar to conformance checking. Conceptually, we use the execution paths discovered in step 1 to perform a sort of symbolic execution over the BPMN model generated by Pragmos, thus identifying discrepancies that can be associated to missing causality relations and/or rewriting loops. Concretely, we implement a procedure where each execution path gets aligned with the model over the modular decomposition tree. The idea works as follows. Each node in the modular decomposition tree is annotated with the set of descendants, that is, a set with the names of activities that can be reached from the root of a given subtree. By testing set inclusion, we can navigate the modular decomposition tree to align each execution path. In our example, when we reach the repeat-like loop, we will find that, for Execution path 1, none of the activities in the loop body ("Complete Retake Exam" and "Grade Retake Exam") is part of such path. That would lead to changing the type loop into a while-like loop. This change would allow a perfect alignment of Execution path 1 with our model.

In the case of activity/block skipping, the alignment process would allow us to identify the subset of activities (or blocks) that must be skipped. Again, the subset of activities (or blocks) to be skipped must form a SESE fragment (well-formedness requirement), add a new activity with a name which will be handled as an invisible activity, and put such a new activity in conflict with all the activities to be skipped, and in causality relation with the transitive predecessor and transitive successor activities that are common to all the activities to be skipped.

\begin{figure}
\begin{tikzpicture}
\node at (0,0) (dfg) {
\begin{tikzpicture}
\node[draw,thick,minimum width=5mm,minimum height=5mm] at (0,0) (a) {a};
\node[draw,thick,minimum width=5mm,minimum height=5mm,right=5mm of a] (b) {b};
\node[draw,thick,minimum width=5mm,minimum height=5mm,right=5mm of b] (c) {c};
\node[draw,thick,minimum width=5mm,minimum height=5mm,right=5mm of c] (d) {d};
\node[draw,thick,minimum width=5mm,minimum height=5mm,below right=6 mm and 10mm of a] (tau) {$\tau$};
\node[draw,thick,circle,minimum width=4mm,left=4mm of a] (start) {}; 
\node[draw,fill=green!20,isosceles triangle,isosceles triangle apex angle=65,minimum size=2mm,inner sep=0] at (start) {}; 
\node[draw,thick,circle,minimum width=4mm,right=4mm of d] (end) {}; 
\node[draw,fill=red!20,minimum size=2mm,inner sep=0] at (end) {};
\draw[thick,->] (start) edge (a) (a) edge (b) (b) edge (c)
    (c) edge (d) (d) edge (end)
    (a) edge [bend right=40] (c) (a) edge [bend left=40] (d.north)
    (b) edge [bend left= 40] (d)
    (a) edge [bend right=20] (tau) (tau) edge  [bend right=20] (d)
    ;
\end{tikzpicture}
};
\node at (5,0) (org) {
\begin{tikzpicture}
\node[draw,thick,minimum width=5mm,minimum height=5mm] at (0,0) (a) {a};
\node[draw,thick,minimum width=5mm,minimum height=5mm,above right=-1mm and 5mm of a] (b) {b};
\node[draw,thick,minimum width=5mm,minimum height=5mm,right=4mm of b] (c) {c};
\node[draw,thick,minimum width=5mm,minimum height=5mm,below right=-1mm and 5mm of c] (d) {d};
\node[draw,thick,minimum width=5mm,minimum height=5mm,below right=-1 mm and 10mm of a] (tau) {$\tau$};

\node[draw,rounded corners,fit=(b) (c)] (bc) {};
\node[draw,rounded corners,fit=(bc) (tau)] (bct) {};
\node[draw,rounded corners,fit=(a) (bct) (d)] (all) {};

\draw[thick,->] (a) edge (bct) (bct) edge (d) (b) edge (c);
\end{tikzpicture}
};
\node at (3,-3) {
\begin{tikzpicture}[rounded corners]
\node[draw,thick,minimum width=5mm,minimum height=5mm] at (0,0) (a) {a};
\node[draw,thick,minimum width=5mm,minimum height=5mm,above right=0mm and 8mm of a] (b) {b};
\node[draw,thick,minimum width=5mm,minimum height=5mm,right=4mm of b] (c) {c};
\node[draw,thick,minimum width=5mm,minimum height=5mm,below right=0mm and 8mm of c] (d) {d};

\node[draw,thick,diamond,right=3mm of a,rounded corners=0] (xor1) {};
\node at (xor1) {$\Xor$};
\node[draw,thick,diamond,left=3mm of d,rounded corners=0] (xor2) {};
\node at (xor2) {$\Xor$};
\node[draw,thick,circle,minimum width=4mm,left=4mm of a] (start) {}; 
\node[draw,ultra thick,circle,minimum width=4mm,right=4mm of d] (end) {};
\draw[thick,->] (start) edge (a) (a) edge (xor1) (xor2) edge (d)
    (b) edge (c) (d) edge (end);
\draw[thick,->,rounded corners=0] (xor1) |- (b);
\draw[thick,->,rounded corners=0] (c) -| (xor2);
\draw[thick,->,rounded corners=0] (xor1) -- ++(0,-6mm) -- ([shift={(0,-6mm)}]xor2.center) -- (xor2);
\end{tikzpicture}
};
\end{tikzpicture}
\caption{}
\end{figure}

We show an example on how to add the invisible activity, used to handle the activity skipping. Let us assume that, initially, we have a model with activities $a$, $b$, $c$ and $d$ in a sequence. The corresponding ordering relations graph is shown in Fig.~\ref{} where we have also added an invisible activity ($\tau$) in conflict with activities $b$ and $c$. Please note that the transitive predecessor (successor) of $b$ and $c$ is $a$ ($d$, respectively). Hence, $a$ will be a predecessor of the invisible activity and $d$ its successor, as shown in Fig.~\ref{}. The modular decomposition tree will now have a model with $b$ and $c$ in causality, which is in conflict with $\tau$. Finally, there is an outer-most module that captures a sequence (i.e., a module with a causality relation). The BPMN model generated from the modular decomposition is shown in Fig.~\ref{}, where we do not materialize the invisible activity, but only a sequence flow to skip activities $b$ and $c$, as expected.

\section{Evaluation}
\label{sec:evaluation}

The ideas presented in the previous section were implemented in Pragmos in Python, with connections to LLMs using the OpenAI API, Google's Gemini API. Thus, Pragmos can be connected to ChatGPT, GPT-OSS, Gemma and Gemini. This way, it is possible to use Pragmos fully with local models (using Gemma and GPT-OSS) as well as with cloud-based services such as ChatGPT and Gemini.

For the evaluation, we used the well-known PET dataset~\cite{BellanADGP22}. This dataset consists of 45 process descriptions, resulting in models of typical sizes. Pragmos generated a business process model for each one of the process descriptions. After a manual verification by the authors, the process models capture appropriately the logic described in the document. The only exception is, probably, the process description ``doc-4.1'' which is very large, and resulted in a model with a large number of activities for step 1. This model exhibited one technical constraint with the current implementation of the modular decomposition.

\section{Conclusions and future work}
\label{sec:conclusions}

In this paper we introduced Pragmos an Agent-based system for process modeling. Different to other existing tools, Pragmos produces a series of process models that reflect several stages of modeling. Each stage is generated from intuitive information generated by an LLM, which can be used by Analysts to track, verify and even refine the models. We contend that using these intuitive artifacts is important to allow analysts enter the inherent co-design cycle. Although the resulting models are similar to the ones produced by other tools, such tools use highly technical internal artifacts (e.g., generated Python code) which hinders the direct intervention of analysts.

We evaluated Pragmos with the PET dataset. Our tool was able to produce correct (sound) process models for most of the cases. Notable exceptions were some of the processes descriptions which were particularly large in extent. Although Pragmos provides already support for fragment abstraction, the descriptions were extremely large such that the maximum context length was exceeded. This is one of the avenues we need to explore in future research. On the other hand, we need to validate the usefulness of Pragmos with an empirical study with analysts.

\subsubsection{\ackname} 
This work has received funding from the Swiss National Science Foundation under Grant No. IZSTZ0\_208497 (ProAmbitIon project).

%
%
%
 \bibliographystyle{splncs04}
 \bibliography{literature}
\end{document}